\documentstyle[epsfig,12pt,dina4]{article}

\begin{document}

\baselineskip=18pt      %contribution

\begin{center}

{\Large\bf
Source Temperatures and Sizes
\vspace{0.2cm}

in Central Collisions}

%(draft \today)
\vspace*{0.6cm}

\normalsize                % Zeilenabstand zu schreiben

C. Schwarz$^{(1)}$ for the ALADIN collaboration:

\vspace*{0.1cm}

R.~Bassini,$^{(2)}$
M.~Begemann-Blaich,$^{(1)}$
A.S.~Botvina,$^{(3)}$\footnote{
Present Address: Bereich Theoretische Physik, Hahn-Meitner-Institut,
D-14109 Berlin, Germany}
S.~Fritz,$^{(1)}$
S.J.~Gaff,$^{(4)}$
C.~Gro\ss,$^{(1)}$
G.~Imm\'{e},$^{(5)}$
I.~Iori,$^{(2)}$
U.~Kleinevo\ss,$^{(1)}$
G.J.~Kunde,$^{(4)}$
W.D.~Kunze,$^{(1)}$
U.~Lynen,$^{(1)}$
V.~Maddalena,$^{(5)}$                   %Valentina
M.~Mahi,$^{(1)}$
T.~M\"ohlenkamp,$^{(6)}$
A.~Moroni,$^{(2)}$
W.F.J.~M\"uller,$^{(1)}$
C.~Nociforo,$^{(5)}$                    %Chiara
B.~Ocker,$^{(7)}$
T.~Odeh,$^{(1)}$
F.~Petruzzelli,$^{(2)}$
J.~Pochodzalla,$^{(1)}$\footnote{
Present address: Max-Planck-Institut f\"ur Kernphysik,
D-69117 Heidelberg, Germany}
G.~Raciti,$^{(5)}$
G.~Riccobene,$^{(5)}$                   %Giorgio
F.P.~Romano,$^{(5)}$                    %Paolo
Th.~Rubehn,$^{(1)}$\footnote{
Present address: Nuclear Science Division, Lawrence Berkeley Laboratory,
Berkeley, CA 94720, USA}
A.~Saija,$^{(5)}$                       %Andrea
M.~Schnittker,$^{(1)}$
A.~Sch\"uttauf,$^{(7)}$
W.~Seidel,$^{(6)}$
V.~Serfling,$^{(1)}$
C.~Sfienti,$^{(5)}$                     %Titti
W.~Trautmann,$^{(1)}$
A.~Trzcinski,$^{(8)}$
G.~Verde,$^{(5)}$
A.~W\"orner,$^{(1)}$
Hongfei~Xi,$^{(1)}$\footnote{
Present address: NSCL, Michigan State University,
East Lansing, MI 48824, USA }
and B.~Zwieglinski$^{(8)}$

\vspace{0.5cm}

$^{(1)}$Gesellschaft  f\"ur  Schwerionenforschung, D-64291 Darmstadt,
Germany\\
$^{(2)}$Istituto di Scienze Fisiche dell' Universit\`{a}
and I.N.F.N., I-20133 Milano, Italy\\
$^{(3)}$Institute for Nuclear Research,
Russian Academy of Sciences, 117312 Moscow , Russia\\
$^{(4)}$Department of Physics and
Astronomy and National Superconducting Cyclotron Laboratory,
Michigan State University, East Lansing, MI 48824, USA\\
$^{(5)}$Dipartimento di Fisica dell' Universit\`{a}
and I.N.F.N.,
I-95129 Catania, Italy\\
$^{(6)}$Forschungszentrum Rossendorf, D-01314 Dresden, Germany\\
$^{(7)}$Institut f\"ur Kernphysik,
Universit\"at Frankfurt, D-60486 Frankfurt, Germany\\
$^{(8)}$Soltan Institute for Nuclear Studies,
00-681 Warsaw, Hoza 69, Poland

\vspace{0.6cm}

{ABSTRACT}
\end{center}

\normalsize                % Zeilenabstand zu schreiben

For midrapidity fragments from central 50-200 AMeV Au+Au collisions
temperatures from double ratios of isotopic yields were compared with
temperatures from particle unbound states. Temperatures from particle
unbound states with $T\simeq 4-5MeV$ show with increasing beam energy an
increasing difference to temperatures from double ratios of isotopic yields,
which increase from $T\simeq 5MeV$ to $T\simeq 12MeV$. The lower
temperatures extracted from particle unstable states can be explained by
increasing cooling of the decaying system due to expansion. This expansion
is driven by the radial flow, and freeze out of particle unstable states
might depend on the dynamics of the expanding system. Source sizes from
pp-correlation functions were found to be 9 to 11 fm.

\newpage

\begin{figure}[htb]
 \centerline{\epsfig{file=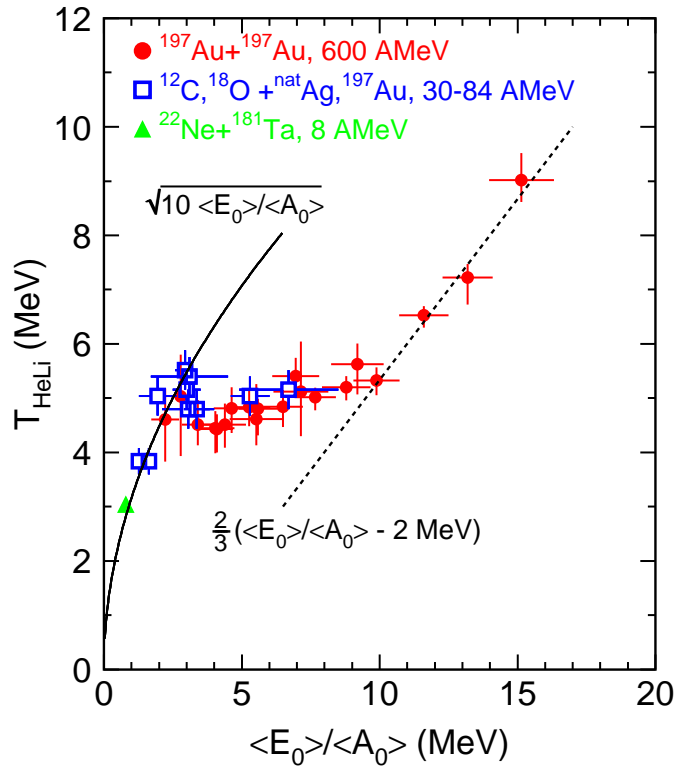,height=12cm}}
%bbllx=0,bblly=0,bburx=567,bbury=567,angle=0}}
  \caption[]{Caloric curve from Ref. \protect{\cite{Pochodzalla95}}}
  \label{Caloric_curve}
\end{figure}

Highly excited nuclear matter with excitation energies above the binding
energy is formed in collisions between heavy nuclei. Projectile spectators
from 600 AMeV Au + Au collisions \cite{Pochodzalla95,Papp,Wolfgang}
show with increasing
excitation energy per nucleon after a temperature rise according to a Fermi
liquid a temperature plateau of about $T\simeq 5~MeV$
(Fig. \ref{Caloric_curve}). Farther excitation
shows up as a linear temperature increase in accordance with a gas of free
nucleons. This relation commonly referred to as caloric curve was
interpreted by the authors as reminiscent of a nuclear liquid-gas phase
transition \cite{Pochodzalla95}.
%%%%%%%%%%%%%%%%%%%%%%%%%%%%%%%%%%%%%%%%%%%%%%%%%%%%%%%%%%%%%%%%%%%%%%%%%
%%%%%%%%%%%%%%%%%%%%%%%%%%%%%%%%%%%%%%%%%%%%%%%%%%%%%%%%%%%%%%%%%%%%%%%%%
The above temperatures were deduced from double ratios of
isotopic fragment yields \cite{Albergo85}. The underlying assumptions of thermal
and chemical equilibrium for
the mass action law for this temperature measurement are also questioned
taking into account interaction between particles leading to Mott
transitions \cite{Roepke}.The following debate about the order of the phase
transition gained a great deal by alternative interpretations like
sequential feeding \cite{Tsang96,Gulminelli971} and
Coulomb instabilities \cite{Natowitz95}
of systems, where the mass decreases with the excitation energies. It is
therefore of interest to extract the caloric curve of participant matter
with nearly constant mass and increasing excitation energy. To shed light on
this problems it is helpful to compare the chemical temperatures with
temperatures deduced from the population of particle unstable states which
are well established for temperatures up to $T=6~MeV$ \cite{Schwarz93,Kunde91}. Therefore, we performed an experiment to measure both
temperatures for midrapidity fragments in Au+Au collisions.

\begin{figure}[b]
 \centerline{\epsfig{file=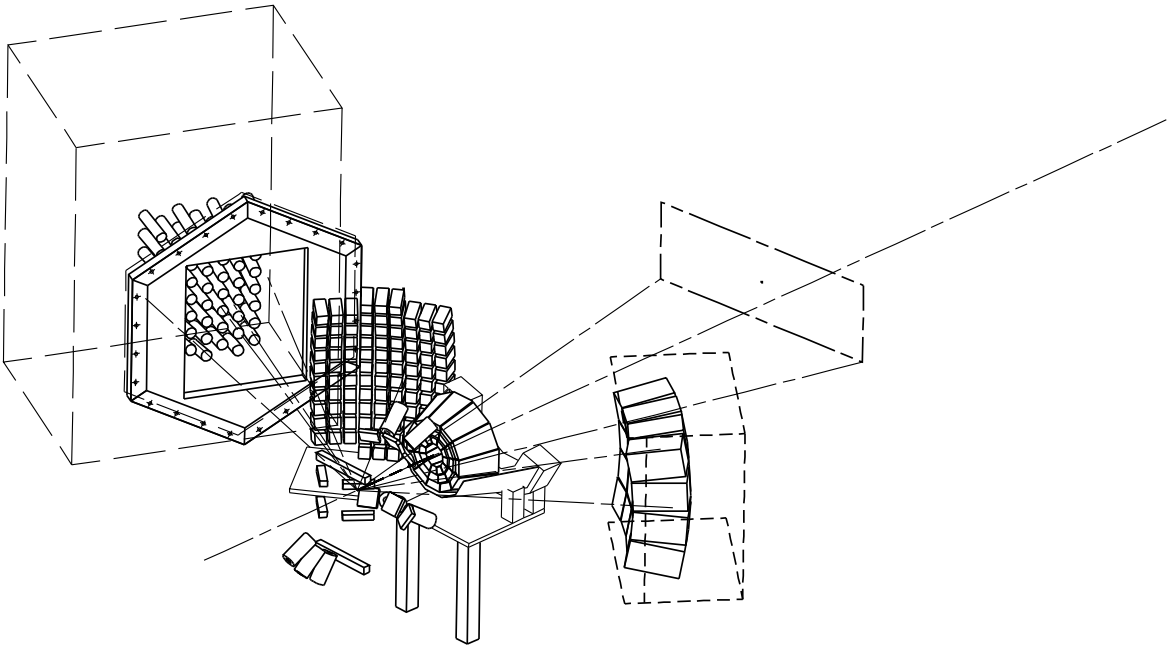,height=10cm,
bbllx=70,bblly=306,bburx=593,bbury=629,angle=0}}
  \caption[]{Setup with 3 hodoscopes and the zero degree hodoscope. The beam
comes from the left. The GSI hodoscope on the right side and the Catania hodoscope
on the left side measured two-particle correlation-functions. On the same side
the MSU hodoscope was setup at larger angles. The Zero Degree Outer
hodoscope (ZDO) arround the beam axis served together with 48 Si-strip
detectors at angles, \mbox{$\theta \approx 40^o$} for the impact parameter
selection. Isotopic resolved fragment spectra were measured with 7 telescopes
arround the target.}
  \label{Setup}
\end{figure}

The experiment was performed using $^{197}Au$ beams of 50, 100, 150, and 200
AMeV, respectively, extracted from the heavy-ion synchrotron SIS of the GSI
facility:

Targets with areal densities of 75 $mg/cm^2$ were irradiated with beam
intensities of \mbox{$10^6~s^{-1}$}. A set of seven telescopes,shown in Fig.
\ref{Setup} consisting of 50,
300, 1000 $\mu m$ Si-surface barrier detectors followed by an 4 cm long
CsI(Tl) scintillator with photodiode readout were used to measure
%%%%%%%%%%%%%%%%%%%%%%%%%%%%%%%%%%%%%%%%%%%%%%%%%%%%%%%%%%%%%%%%%%%%%%%%%
%%%%%%%%%%%%%%%%%%%%%%%%%%%%%%%%%%%%%%%%%%%%%%%%%%%%%%%%%%%%%%%%%%%%%%%%%
isotopically resolved fragment spectra. Four telescopes measured at $\theta
_{lab}=40^{\circ }$ particle yields from the midrapidity source, the other
telescopes were located at $\theta _{lab}=110^{\circ }$, $130^{\circ }$, and 
$150^{\circ }$. Permanent magnets, placed next to the entrance collimator of
each telescope, deflected electrons produced in the target. Isotopic
resolution was achieved up to carbon fragments; the energy thresholds for p,
d, t, $^3$He, $^4$He, $^6$Li, $^7$Li were 2.1, 2.7, 3.1, 6.8, 7.4, 13.9, and 14.4 AMeV,
respectively.
Three hodoscopes built at GSI, LNS Catania, and Michigan State University
and consisting, in total, of 216 Si-CsI(Tl) telescopes were setup at
distances between 0.6 m and 1.1 m from the target.
%%%%%%%%%%%%%%%%%%%%%%%%%%%%%%%%%%%%%%%%%%%%%%%%%%%%%%%%%%%%%%%%%%%%%%%%%
\begin{figure}[b]
\centerline{\epsfig{file=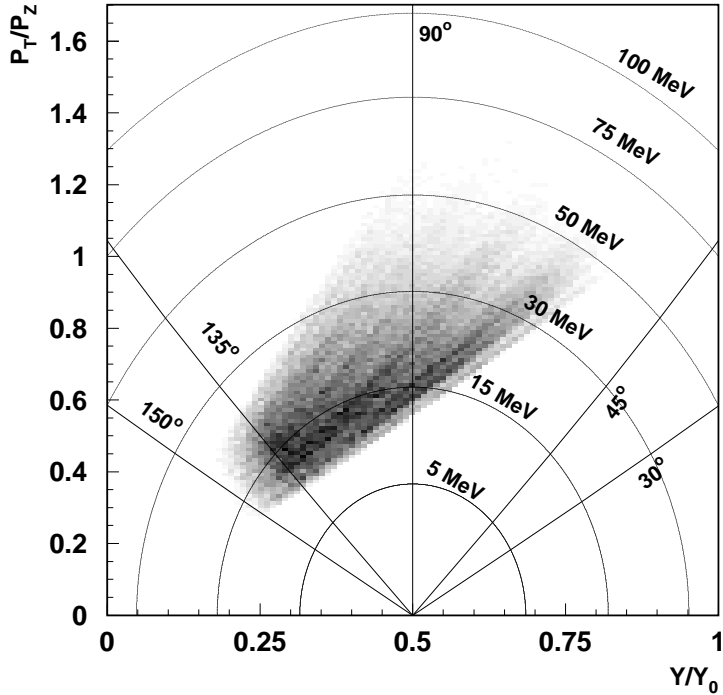,height=10cm}}
%bbllx=70,bblly=306,bburx=593,bbury=629,angle=0}}

\caption{Accpetance of the hodoscopes for detected p-$\alpha $ pairs at 150
AMeV beam energy. At all incident energies the fragments were detected at $%
\theta _{CMS}\approx 90^{\circ }.$ }
\label{accept}
\end{figure}
%%%%%%%%%%%%%%%%%%%%%%%%%%%%%%%%%%%%%%%%%%%%%%%%%%%%%%%%%%%%%%%%%%%%%%%%%
The GSI hodoscope which
consisted of 64 elements and the Catania hodoscope which consisted of 96
elements covered an angular range between $\theta _{lab}=24^{\circ
}-59^{\circ }$.
The correlation data obtained with these
two hodoscopes and the telescopes are the subject of this
paper and consist of an 300 $\mu m$ Si-detector followed by a 6 cm long
CsI(Tl) scintillator with photodiode readout. For the Catania hodoscope,
each telescope had a solid angle of = 2.95 msr, and the angular spacing
between adjacent telescopes was $\Delta \theta =0.22^{\circ }$ . For the GSI
hodoscopes, each 4 detectors were combined in a group with an angular
spacing of $\Delta \theta < 0.01^{\circ }$ within and $\Delta \theta =$ $%
2.5^{\circ }$ between the groups of detectors. The covered solid angle of
each detector was $\Delta \Omega =1.35~msr$. Energy calibration for
individual detectors of the hodoscopes were obtained by employing energies
of fragments just stopped in the dE-Silicon counters and the CsI(Tl)
crystals. This resulted in calibrations with accuracies of about 2 \% for
protons and 3 \% for Lithium. These calibration accuracies were assessed by
comparing peak widths of experimental and calculated correlation functions.
In addition, 36 phoswich detectors and 48 Si-strip detectors from $%
6.5^{\circ }$ to $40^{\circ }$ increased the solid angle and the granularity
of the detector system and allowed impact parameter selection.

The angular range of our hodoscopes were chosen around $\theta
_{cm}=90^{\circ }$ in order to minimize contributions from target like and
projectile like sources. The acceptance for the GSI and the Catania
hodoscope is shown in Fig. \ref{accept} for $^5Li$, reconstructed from
coincident measured proton-alpha pairs.
Charged particle multiplicities served as impact parameter filter,
and we selected the most central collisions (9\% of yield) in accordance
with an reduced impact parameter range of $\widehat{b}=0.3$.

%%%%%%%%%%%%%%%%%%%%%%%%%%%%%%%%%%%%%%%%%%%%%%%%%%%%%%%%%%%%%%%%%%%%%%%%%
\begin{figure}[t]
 \centerline{\epsfig{file=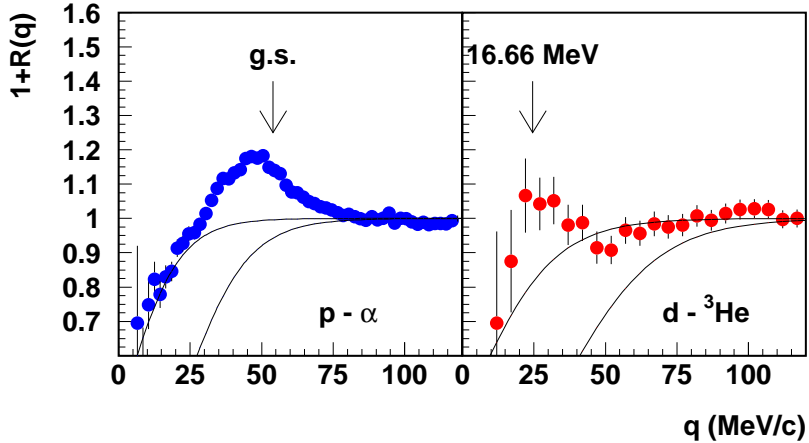,height=7cm}}
%bbllx=0,bblly=0,bburx=567,bbury=567,angle=0}}
  \caption[]{Correlation functions coincident p-$\alpha$ pairs (left panel)
and d-$^3$He pairs (right panel) at 150 AMeV beam energy.
The lines denote extrem assumptions about the
background correlation function which were guided by scaled correlation
functions of nonresonant states.  The nonresonant correlation functions were
assumed to scale with the relative velocity between the pairs.
with }
  \label{Li5}
\end{figure}
%%%%%%%%%%%%%%%%%%%%%%%%%%%%%%%%%%%%%%%%%%%%%%%%%%%%%%%%%%%%%%%%%%%%%%%%%
We employed correlation functions in order to determine temperatures from
particle unstable states. For the extraction of the yields we assumed two
extreme assumptions of the background correlation function which determine
our systematical errors. They were constructed with the help of the
correlation functions from fragments without resonant states. For technical
details of the extraction of the temperatures see Ref. \cite{Schwarz93}. We also
used the decay of $^8Be$ as thermometer, the results for all
incident energies are shown in Tab. \ref{temptable}. They all agree with an
average apparent temperature of $T\approx 4-5$ MeV. As an example for the decay of
$^5$Li the correlation function of its decay products is shown in
\mbox{Fig. \ref{Li5}}.

\begin{figure}[t]
 \centerline{\epsfig{file=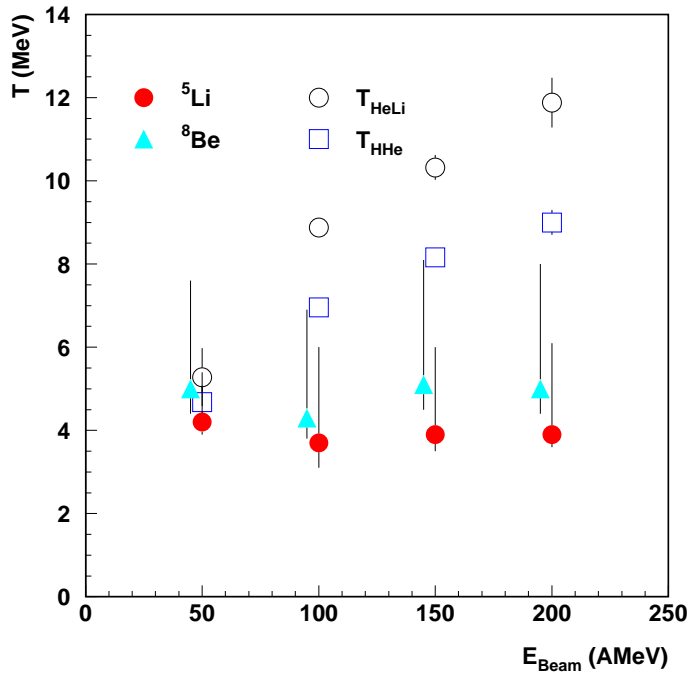,height=10cm}}
%bbllx=0,bblly=0,bburx=567,bbury=567,angle=0}}
\caption{Temperatures from double ratios of isotopic yields (open symbols)
compared to temperatures from particle unstable states (closed symbols).
}
\label{temperature}
\end{figure}

Chemical temperatures, extracted from double ratios of $%
(^3He/^4He)/(^6Li/^7Li)$ and $(d/t)/(^3He/^4He)$ \cite{Pochodzalla95}, for
the Si-telescope at a polar angle of $\theta _{lab}=40^{\circ }$ , are shown
in Fig. \ref{temperature} for central collisions. For deuterons and tritons
a correction for the limited acceptance of particles traversing through the
detector was done with the help of a moving-source parametrization. Low energy
thresholds were low enough to observe the maximum in the energy spectra.
By estimating the excitation energy of the reaction zone it is possible to
compare these temperatures with those of the caloric curve for spectator matter
in Fig. \ref{Caloric_curve}.
Using as excitation energy the maximum possible excitation energy
($E_{beam}/4$) and subtracting the known values of collective motion (radial
flow)
\cite{Poggi95} the four datapoints resemble the second rise in the
caloric curve.
%%%%%%%%%%%%%%%%%%%%%%%%%%%%%%%%%%%%%%%%%%%%%%%%%%%%%%%%%%%%%%%%%%%%%%%%%
\begin{table}[t]
\centerline{
\begin{tabular}{|r|c|r|c|r|}
\hline
$E_{Beam}$ (AMeV) & \multicolumn{2}{|c|}{$T_{excited}$ (MeV)} & 
\multicolumn{2}{|c|}{$T_{double~rat.}$ (MeV)}\\ \hline
$50 MeV/A$ & $^5Li$ & $4.2^{+1.2}_{-0.3}$ & $(^6Li/^7Li)/(^3He/^4He) $ & $%
5.3 \pm 0.8$ \\ 
& $^8Be$ & $5.0^{+2.6}_{-0.6}$ & $(^6Li/^7Li)/(^2H/^3H)$ & $4.7 \pm 0.5$ \\ 
\hline
$100 MeV/A$ & $^5Li$ & $3.7^{+2.3}_{-0.6}$ & $(^6Li/^7Li)/(^3He/^4He) $ & $%
8.9 \pm 0.1$ \\ 
& $^8Be$ & $4.3^{+2.6}_{-0.5}$ & $(^6Li/^7Li)/(^2H/^3H)$ & $7.0 \pm 0.1$ \\ 
\hline
$150 MeV/A$ & $^5Li$ & $3.9^{+2.1}_{-0.4}$ & $(^6Li/^7Li)/(^3He/^4He) $ & $%
10.3 \pm 0.4$ \\ 
& $^8Be$ & $5.1^{+3.0}_{-0.6}$ & $(^6Li/^7Li)/(^2H/^3H)$ & $8.2 \pm 0.2$  \\ 
\hline
$200 MeV/A$ & $^5Li$ & $3.9^{+2.2}_{-0.3}$ & $(^6Li/^7Li)/(^3He/^4He) $ & $%
11.9 \pm 0.4$ \\ 
& $^8Be$ & $5.0^{+3.0}_{-0.6}$ & $(^6Li/^7Li)/(^2H/^3H)$ & $9.0 \pm 0.4$ \\ 
\hline
\end{tabular}
}
\caption{Temperatures for central collision ($\hat{b}\le 0.3$) extracted from
particle unstable states ($T_{excited}$) and from double ratios of isotopic
yields ($T_{doublerat.}$) with their systematical errors. For statistical
reasons no impact parameter cut was applied to the 50 AMeV data}
\label{temptable}
\end{table}
%%%%%%%%%%%%%%%%%%%%%%%%%%%%%%%%%%%%%%%%%%%%%%%%%%%%%%%%%%%%%%%%%%%%%%%%%

It is surprising that with increasing beam energy there is a growing
disagreement between chemical temperatures from double ratios of fragment
yields and temperatures extracted from particle unbound states. While for 50
AMeV beam energy both temperatures agree within the error bars, the chemical
temperatures at 200 AMeV beam energy are higher by a factor of 4 compared to
temperatures from excited states. A known effect, the sequential feeding,
which alters the measured temperatures, is still under investigation
\cite{Tsang96,Gulminelli971}.
For the chemical temperatures, however, we have applied a
sequential feeding correction of a factor of 1.2 to the temperatures, which
we determined from model comparisons \cite{Pochodzalla95}. For temperatures
from particle unbound states many investigations of sequential feeding were
performed (see eg. \cite{Chen88}). Especially, temperatures from $^5Li$
and $^8Be$ were shown to be quite robust against sequential feeding for
temperatures $T<5MeV$ .
%%%%%%%%%%%%%%%%%%%%%%%%%%%%%%%%%%%%%%%%%%%%%%%%%%%%%%%%%%%%%%%%%%%%%%%%%
%%%%%%%%%%%%%%%%%%%%%%%%%%%%%%%%%%%%%%%%%%%%%%%%%%%%%%%%%%%%%%%%%%%%%%%%%

Taking the temperature difference of both thermometers for true, the different
freeze out conditions for both thermometers must be different. In the
initial stage of the reaction the density will be high and fragments will
not yet have their identity. Below a certain density fragments will be
formed and destroyed by the surrounding hot nuclear medium. When the fragments
have reached their final identity (chemical equilibrium is achieved), the
isotopic ratios reflect temperatures which we have presented in this work. A
further step in time order is the formation of excited states. If one
assumes that the excited states are quite fragile objects in the surrounding
gas of nucleons they will be destroyed and formed later in time as compared
to compact particle stable fragments. Therefore, the temperatures from excited states
might be lower because of the cooling of the reaction zone. Indeed, NMD
calculations \cite{Barz96} show for central Au+Au collisions at 150 AMeV
that fragments already formed suffer for time intervals up to
30-50 fm/c collisions with the surrounding environment.

\begin{figure}[h]
\centerline{
\epsfig{file=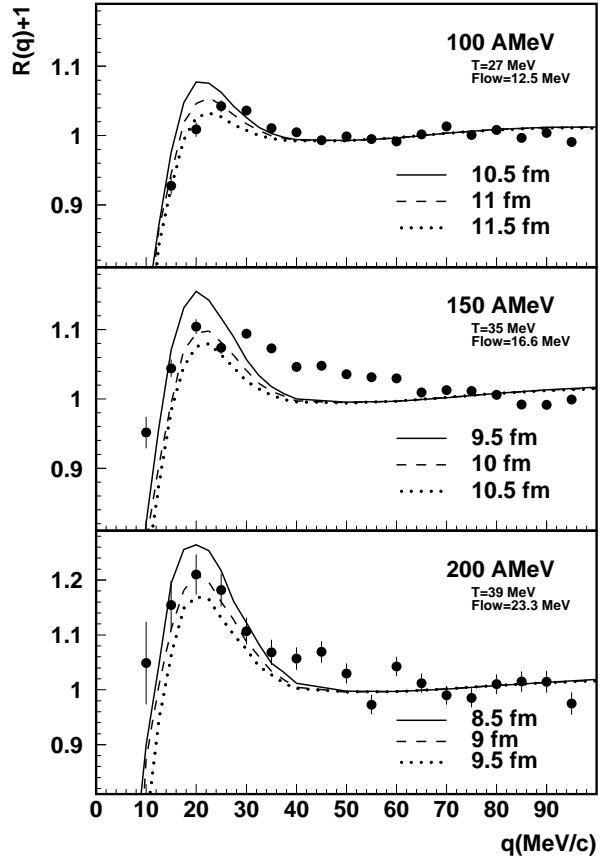,width=8.2cm,bbllx=0,bblly=0,bburx=524,bbury=788}
}
\caption{{\protect Proton-Proton correlation functions are shown for
central collision at the indicated incident energies. Theoretical predictions
(lines) indicate a shrinking apparant source size. }}
\label{pp}
\end{figure}

The expanding source scenario that evolves with
increasing incident energies also
influences the proton-proton correlation function,
from which information about
the space-time behaviour of the source may be obtained \cite{Xi95}.
The correlation functions for proton pairs from reactions
at 100, 150, and 200 AMeV incident energies are shown in Fig. \ref{pp} (data
points). From
QSM \cite{QSM} calculations, with parameters selected to describe measured
yields, the fractions of protons emitted from long lived resonances were
estimated to be 0.21, 0.29, and 0.31 for 100, 150, and 200 AMeV, respectively.
The data were corrected for this effect and source sizes were
deduced by comparing these correlation functions with predictions of the
Koonin-Pratt formalism (lines in Fig. \ref{pp}). A Monte Carlo generated
source with flow values from Ref. \cite{Poggi95}, zero lifetime,
and with slope parameters for the energy
spectra \cite{Poggi95} served as input for
this theoretical description.
The simulated protons have passed through the experimental filter.
%%%%%%%%%%%%%%%
%%%%%%%%%%%%%%%%%
With increasing incident energy the apparent source size decreases from
11 fm to 9 fm (hard sphere radius).
These values are larger than the radius of the
combined system of two Au nuclei at
groundstate density. The systematical errors resulting from
different methods of normalization are estimated to be about 0.5 fm.

In conclusion, we found for increasing beam energies of 50, 100, 150, and
200 AMeV Au+Au, for midrapidity fragments an increasing deviation between
temperatures from double ratios of isotopic yields and temperatures
extracted from excited state populations. This deviation can be explained by
different freeze out conditions and different cooling of the nucleus. Then,
it allows to investigate freeze out conditions for stable fragments
and excited states. 
Source sizes from proton-proton correlation functions
indicate a shrinking source size with increasing beam energy of 11 to 9 fm.

\vspace{0.2cm}

\renewcommand{\baselinestretch}{0.85}
\Large                % 1. Huerde:
                      % Erst groesser, um dann normal mit kleinerem
\normalsize                % Zeilenabstand zu schreiben

\end{document}